\documentstyle[11pt,amsfonts]{article}

\newcommand{\be}{\begin{equation}}
\newcommand{\ee}{\end{equation}}
\newcommand{\bea}{\begin{array}}
\newcommand{\ea}{\end{array}}
\newcommand{\beqa}{\begin{eqnarray}}
\newcommand{\eeqa}{\end{eqnarray}}
\newcommand{\bean}{\begin{eqnarray*}}
\newcommand{\eean}{\end{eqnarray*}}

\def\up#1{\leavevmode \raise.16ex\hbox{#1}}

\newcommand{\gapproxeq}{\lower
 .7ex\hbox{$\;\stackrel{\textstyle >}{\sim}\;$}}
\newcommand{\lapproxeq}{\lower .7ex\hbox{$\;\stackrel
{\textstyle <}{\sim}\;$}}

\newcounter{appendice}

\def\thebibliography#1{{\bf REFERENCES\markboth
 {REFERENCES}{REFERENCES}}\list
 {[\arabic{enumi}]}{\settowidth\labelwidth{[#1]}\leftmargin\labelwidth
 \advance\leftmargin\labelsep
 \usecounter{enumi}}
 \def\newblock{\hskip .11em plus .33em minus -.07em}
 \sloppy
 \sfcode`\.=1000\relax}

\begin{document}
\begin{flushright}
 SU-4252-791\\
\end{flushright}
\centerline{ \LARGE Space-Time Noncommutativity from Particle Mechanics } 

\vskip 2cm

\centerline{ {\sc    A. Pinzul$^{a}$ and A. Stern$^{b}$ }  }

\vskip 1cm
\begin{center}
{\it a)  Department of Physics, Syracuse University,\\ Syracuse, 
New York 13244-1130,  USA\\}
{\it b) Department of Physics, University of Alabama,\\
Tuscaloosa, Alabama 35487, USA}

\end{center}

\vskip 2cm

\vspace*{5mm}

\normalsize
\centerline{\bf ABSTRACT} 
We exploit the  reparametrization symmetry of a relativistic free
particle to impose a gauge condition which upon quantization implies
space-time noncommutativity.   We show that there is an algebraic map
from this gauge back to the standard `commuting'  gauge.
Therefore the  Poisson
algebra, and the resulting  quantum theory, are identical in the  two gauges.  The
only difference is in the interpretation of space-time coordinates.
The procedure is repeated for the case of a coupling with a
constant electromagnetic field, where the reparametrization symmetry is preserved.
For more  arbitrary interactions,
  we show that standard dynamical
 system can be  rendered noncommutative in space and time by a simple change of
 variables.

\vspace*{5mm}

\newpage
\scrollmode

\section{Introduction}

Issues  concerning the loss of unitarity have been raised in the
context of field theories
with space-time noncommutativity, despite the work of Doplicher, Fredenhagen and Roberts\cite{Doplicher:1994tu} to
the contrary.  In this regard, it might be useful to
examine space-time noncommutativity in a simpler setting.
In the context of quantum mechanics, space-time noncommutativity can
be introduced in a trivial manner.  Say that ${\bf x}^i $ and ${\bf
  p}_i $ are the position and momentum operators for a particle
satisfying
\be [{\bf x}^i,{\bf p}_j ] =i\delta_{ij}\;,\label{ccrqp} \ee
and evolution in some  variable $\tau$ is generated by Hamiltonian
${\bf H}$.  We usually  call $\tau$ the `time'.  Alternatively, there
have been attempts to make the time, like the spatial coordinate, be
associated with a quantum operator.\cite{Sorkin}  This allows for the exotic
possibility of having the space and time coordinates  be noncommuting.  A
trivial way to achieve this is to declare the `time operator' to be
\be  {\bf x}^0 = \tau-\theta^{0i}{\bf p}_i\;, \label{bfxz}\ee
where $ \theta^{0i}$ are constants.  When  $ \theta^{0i}\rightarrow 0$
one recovers the commutative time, while for  $ \theta^{0i}\ne 0$,
 \be [{\bf x}^0,{\bf x}^i]= i\theta^{0i}\label{stcrs}\ee
Similar redefinitions can be done to introduce noncommutativity among
only spatial coordinates \cite{Duval:2000xr},\cite{Nair:2000ii}.
Balachandran, et. al.\cite{bal} have developed a quantum theory based on
 commutation relations (\ref{stcrs}).

  In this note we show that in theories with time
  reparametrization symmetry, space-time noncommutativity is simply 
  a gauge choice.   We consider  familiar examples in particle mechanics.  In section 2 we
  re-examine the 
relativistic free  particle.  The action is reparametrization invariant with
respect to the  parameter labeling the position along the world line.   By choosing a nonstandard gauge
condition we can obtain Dirac brackets corresponding to the classical
  analogue of  (\ref{stcrs}).  The situation  resembles the derivation of
  spatial noncommutativity for a charged particle in a strong magnetic
  field.\cite{Szabo}  As the
classical   physics cannot depend on the
gauge choice, this theory should be equivalent to the 
theory expressed in the standard gauge, where the parameter is
identified with the time coordinate.  This equivalence can be made
explicit by displaying a simple algebraic  map between the two
theories.  The time component of it is given by (\ref{bfxz}).
Introducing interactions will in general  spoil the  reparametization
symmetry present for the free particle.  An exceptional case is the coupling
to an electromagnetic background.  We consider the case of a constant
  electromagnetic background in section
  3.  As before we show that there is a gauge
condition which leads to  (\ref{stcrs}) upon upon quantization. 
Also as before,  the
 noncommuting space-time coordinates can be
obtained by  applying a coordinate transformation from the
 standard gauge. 

 In both of the above mentioned examples the only difference between the different  gauges is
  what one chooses to call the `time'.  In the above ${\bf x}^0$ and $
  \tau$ represent a `noncommutative' and `commutative time', respectively. Furthermore, time as measured
  by ${\bf x}^0$ and $ \tau$ runs at the same rate (at least classically).  This is evident for the free particle,
  using (\ref{bfxz}), since the momentum is conserved, and hence
  $\frac{d{\bf x}^0}{d\tau}=1$.  It is also true in the case of
  interactions with a constant electromagnetic field provided one interprets
  ${\bf p}_i$ in (\ref{bfxz}) as the conserved momenta.  On the
  other hand,  $\frac{d{\bf x}^0}{d\tau}\ne1$ for arbitrary interactions, which we
  briefly consider in section 4.   Furthermore, one has the
  possibility of  $\frac{d{\bf x}^0}{d\tau}<0$ implying a time
  reversal upon mapping `time'  $\tau$ to `time' ${\bf x}^0$ using 
  (\ref{bfxz}). 

\section{Free Particle}
\setcounter{equation}{0}

We start with the standard reparametrization invariant action for a
relativistic  free
particle in $d+1$ dimensions
\be
S_0 =-m\int d\tau \;\sqrt{- \dot x^2}\;, \label{fpa}
\ee 
with $x^\mu$, $\mu=0,1,...d$ being the space-time coordinate,   the dot denoting differentiation
with the  affine time $\tau$, and metric $\eta=$diag$(-1,1,...,1)$.
From the equations of motion, the momenta \be p_\mu = \frac{m\dot
  x_\mu}{\sqrt{- \dot x^2}} \label{smlpmu}\ee
  are conserved.  
In the gauge invariant formulation of the theory, they are canonically
conjugate to the space-time coordinates, 
\be 
\{x^\mu,p_\nu\} =\delta^\mu_\nu \qquad \{x^\mu,x^\nu\} =
\{p_\mu,p_\nu\} =0\label{cpb}\;, \ee
and are subject to  the mass shell condition
\be
\phi_1 =p^2 + m^2 \approx 0\;,\label{mschl}
\ee  where $\approx$   indicates equality  in weak sense. 
$\phi_1$ generates gauge motion on the phase space associated with
reparametrizations of the parameter $\tau$.   The Poincar\'e symmetry
is generated by $p_\mu$ and $j_{\mu\nu}=x_\mu p_\nu - x_\nu p_\mu $.

The gauge symmetry can be fixed by imposing a gauge condition.  The
standard choice identifies the time coordinate $x^0$ 
with the parameter
 $\tau$.  We instead  impose the following constraint:
\be \phi_2 = x^0 + \theta^{0i} p_i - \tau\;\approx 0,\qquad
i=1,2,...d\;,\label{gfxng} \ee $\theta^{0i}$ being constants.
The constraints (\ref{mschl}) and (\ref{gfxng}) form a  second class
set with
\be \{\phi_1,\phi_2\} = 2 p_0 \label{pboc}\ee
and resulting Dirac brackets\cite{Rht}
\be \{A,B\}_{DB} = \{A,B\} + \frac1{ \{\phi_1,\phi_2\}} \biggl(
\{A,\phi_1\}\{\phi_2,B\} -\{A,\phi_2\}\{\phi_1,B\}\biggr)\label{DB} \ee
  The Dirac bracket
of the spatial coordinates  $x^i$ with the `time' $x^0$ is
\be  \{x^0,x^i\}_{DB} = \theta^{0i}\;, \label{stnc}\ee 
leading to commutation relations (\ref{stcrs}) upon quantization.  The  remaining nonvanishing Dirac brackets are
\beqa
 \{x^i,x^j\}_{DB}& =&\frac1{p_0}\;(\theta^{0i} {p_j} -\theta^{0j}  {p_i})\label{dbxixj}\\
 \{x^i,p_0\}_{DB}& =& \frac {p_i}{p_0}\label{nine} \\
\{x^i,p_j\}_{DB}& =& \delta_{ij}\label{ten} \eeqa 
(\ref{nine}) and (\ref{ten}) are the same as in the standard gauge, while
 (\ref{dbxixj}) implies nontrivial commutation relations among spatial coordinates upon quantization.  
Although $x^0$ gets promoted to a noncommuting operator upon quantization, we can still regard  $\tau$ as a c-number in the
quantum theory.   
Upon imposing $\phi_2=0$ strongly,
 $x^0 + \theta^{0i} p_i$ gets identified
with the parameter
 $\tau$.
By definition $\phi_2$ has zero Dirac bracket with all phase space
variables, and then so does  $x^0 + \theta^{0i} p_i$.  It then is in the
center of the Poisson algebra, and consequently  a c-number in the corresponding quantum
algebra.

The reparametrization symmetry means that the Hamiltonian for the
system is weakly zero, i.e., a linear combination of constraints
$\lambda_a\phi_a$, $a=1,2$, and so the evolution of any function $A$ on phase
space is given by
\be
\dot A\approx \frac{\partial A}{\partial \tau} +
 \lambda_a\{A,\phi_a\}\;, \label{dotA}
\ee   where the dot is a total  $\tau$ derivative.
Imposing that the constraints are preserved in time, i.e., $\dot
\phi_a\approx 0$, fixes the Lagrange multipliers to be 
\be \lambda_1= \frac1{2p_0}\; \frac{\partial \phi_2}{\partial
  \tau}\qquad\quad
\lambda_2= 0 \ee Then if $x^0$ and $p_i$ are presumed to have no explicit
$\tau$ dependence, substitution into  (\ref{dotA}) gives
\be
\dot A\approx \frac{\partial A}{\partial \tau} - \frac1{2p_0}\;
\{A,\phi_1\} \label{dotAtwo}
\ee
Although (\ref{dotAtwo}) correctly reproduces the dynamics, since it
is formulated in terms of Poisson brackets rather than Dirac brackets,
it is not evident how to write it on the reduced phase space in the
form of Hamilton's equations, and consequently the quantum dynamics
in terms of Heisenberg's equations.  
 Alternatively, one can write Hamilton's equations using
 Dirac brackets.   In this approach  the Hamiltonian is not a priori
 determined.  Furthermore, in order to have  $\dot
\phi_a\approx 0$ it becomes necessary for some of the original phase
space variables to have an explicit $\tau$ dependence.  In familiar
examples no such  $\tau$ dependent variables
span  the reduced phase space, as in the case of the free
particle in the $x^0=\tau$ gauge,  where the reduced   phase space is coordinatized by
$x^i$ and $p_i$.  On the other hand, the time coordinate $x^0$ gets an explicit $\tau$
dependence from the gauge condition.  In addition, in the  case of the gauge
(\ref{gfxng}), it is  desirable that $x^0$  appears as  a
degree of freedom in the reduced phase space  since we wish to
recover (\ref{stcrs}) upon quantization.
This is accomplished by using (\ref{gfxng}) to eliminate  one of the
momenta, and so the resulting reduced phase space gets an explict
$\tau$ dependence.
More generally an explicit  $\tau$ dependence may be induced in all of the
original    phase space variables using this approach, as we
illustrate in  section 3. 

Concerning  the free particle in the $x^0=\tau$ gauge it is usual to  choose
\be H=\sqrt{p_ip_i + m^2}\label{hamilt}\ee for  the Hamiltonian,  generating  evolution in
 the parameter $\tau$.   The dynamics follows from
\be \dot A =  \frac{\partial A}{\partial \tau} +
\{A,H\}_{DB} \label{heorps}\ee  The
same  choice can be made for the gauge (\ref{gfxng}).  To recover the
correct equations of motion one assumes that $x^i$
and $p_i$  have no explicit $\tau$ dependence, in either gauge.  As stated above, the same
is not true for `time' coordinate $x^0$.  This follows from the demand that
 $\dot \phi_2=\frac{\partial
  \phi_2}{\partial\tau}=0$, and consequently  
\be \frac{\partial
  x^0}{\partial\tau}=1 \ee As  $\{x^0,H\}_{DB}=0$, it  also follows
that $ \dot x^0 =1$, and as a result the commutative and noncommutative clock, as measured by
 $\tau$ and $x^0$, respectively, run at the same rate.

After the gauge fixing, a one parameter family of  Lorentz  generators
can be constructed
\beqa  \tilde j_{ij}&=&x_i p_j - x_j p_i +\alpha p_0( \theta^{0i} p_j -
\theta^{0j} p_i)\cr
\tilde j_{0i}&=&-x^0 p_i -x_ip_0 -\alpha \theta^{0i}p_0^2-\alpha
\theta^{0j} p_jp_i\;,\label{dfoft}\eeqa $\alpha$ being the parameter.  They satisfy
as usual\beqa
\{\tilde j_{\mu\nu},p_\lambda\}_{DB}& =&\eta_{\mu\lambda}  p_\nu -
\eta_{\nu\lambda}  p_\mu \;,\label{dbjp}\\ \{\tilde j_{\mu\nu},\tilde
j_{\lambda\rho} \}_{DB}& =&\eta_{\mu\lambda}\tilde j_{\nu\rho}
-\eta_{\nu\lambda}\tilde j_{\mu\rho}-\eta_{\mu\rho}\tilde
j_{\nu\lambda} +\eta_{\nu\rho}\tilde j_{\mu\lambda}
\eeqa  From (\ref{dbjp})  the momenta
transform covariantly.  For infinitesimal Lorentz transformations,  \be  \delta_\omega p_\mu =\frac12
\omega^{\lambda\rho} \{p_\mu,\tilde j_{\lambda\rho} \}_{DB}=
-\omega_{\mu\rho}p^\rho \label{top}\ee   Lorentz transformations involve a 
change of gauge, and for that reason  transformations
 of
the space-time coordinates are more subtle.\cite{Rht} $\phi_2$ is not
covariant under Lorentz transformations, since $\theta^{0i}$ are constants.    On the other hand, $\phi_2$,
being in the center of the algebra, has zero Dirac bracket with the
Lorentz generators $\tilde j_{\mu\nu}$.    Therefore Lorentz transformations cannot in general be obtained by
simply taking Dirac brackets with $\tilde j_{\mu\nu}$ as in (\ref{top}).  

As a result of the gauge condition (\ref{gfxng}) we obtained the
 nontrivial Dirac brackets (\ref{stnc}) and  (\ref{dbxixj}) implying
 space-time noncommutativity, as opposed
 to the trivial result for the standard gauge.    However, as was shown
 in \cite{Duval:2000xr},\cite{Nair:2000ii} a simple
 change of variables can remove the noncommutativity.  In this case the
 change is
\beqa  x^i&\rightarrow & q^i = x^i + \theta^{0i} p_0 \label{mpxiqi}
 \\  x^0&\rightarrow & q^0 = x^0 + \theta^{0i} p_i = \tau\;,\label{mpx0q0}\eeqa    (\ref{mpxiqi}) removes the space-space
 noncommutativity implied by  (\ref{dbxixj}), while (\ref{mpx0q0})
 removes the space-time noncommutativity implied by  (\ref{stnc}).
 (\ref{mpx0q0}) also means that the coordinates $q^\mu$ satisfy the
 standard gauge $q^0 = \tau$, and it agrees with (\ref{bfxz}). 
 The only remaining  non zero brackets are
\beqa
 \{q^i,p_0\}_{DB}& =& \frac {p_i}{p_0}\cr
\{q^i,p_j\}_{DB}& =& \delta_{ij}\;,\label{stgg} \eeqa  which agrees
 with the Dirac brackets of the standard gauge.   The free particle
 Hamiltonian is of course unaffected by the coordinate change.  So the only difference
between the two gauges is  the interpretation of the space-time coordinates
 appearing in the free particle action.  Both gauges give rise to an
 identical Poisson structure and dynamics (if we choose $H$ to be the
 same in both gauges), and thus lead to identical
 quantum systems.   Concerning the Lorentz generators, if one sets $\alpha$ in  (\ref{dfoft}) equal to
 one they  have the usual form  \be\tilde
 j_{\mu\nu}=q_\mu p_\nu - q_\nu p_\mu \;,\qquad \alpha=1\ee
As shown in \cite{Rht}, Lorentz transformations of  the space-time
 coordinates $q^\mu$ can be written in a simple form:
  \be \delta_\omega q^\mu=\frac12
\omega^{\lambda\rho} \{q^\mu,\tilde j_{\lambda\rho} \}_{DB}\;- \; \dot q^\mu
\delta \tau \;  = \;-\omega^{\mu\nu}q_\nu  \;\label{btox}\ee 
The subtraction is necessary because the change of  gauge  generated by Lorentz transformations
 corresponds to a shift $\delta \tau$ in $\tau$, while $\tilde
 j_{\mu\nu}$ has zero Dirac bracket with the gauge condition
 $q^0-\tau=0$.  The analogous time derivative term is
absent in the transformation of momentum (\ref{top}) by the equations of motion.
By putting $\mu=0$ in (\ref{btox}),
$ \dot q^0\delta \tau = \omega^{0\mu}q_\mu$, while for $\mu=i$ we then
 get
\be \frac12
\omega^{\lambda\rho} \{q^i,\tilde j_{\lambda\rho} \}_{DB}\;=
 \;\biggl(\frac{ \dot q^i}{ \dot q^0}\; \omega^{0\mu}-\omega^{i\mu}\biggr)q_\mu\;,
 \ee  which is identically satisfied after using the equations of motion.

\section{ Constant Electromagnetic  Field}
\setcounter{equation}{0}

Interactions with an electromagnetic background don't spoil the time reparametrization symmetry
which was present for the relativistic free particle.  In this case a
gauge condition can be imposed which again leads to space-time
noncommutativity upon quantization.   Here we specialize to
 a constant electromagnetic field.  The interaction
term to be added to $S_0$ is then
\be
S_F =-\frac{1}{2}\int d\tau \;
    F_{\mu\nu} x^\mu \dot x^\nu \;\;,\label{cfcplng}
\ee where $F_{\mu\nu}$ is a constant field strength tensor.
The usual equations of motion 
\be \dot p_\mu =-F_{\mu\nu}\dot x^\nu\;,  \label{cpivemf}\ee where $p_\mu$ are
given in (\ref{smlpmu}),  follow from varying $x^\mu$ in  the combined
action $S=S_0 + S_F$.  They  state that
 \be
P_\mu= p_\mu + F_{\mu\nu} x^\nu \label{capP} \ee are constants of the
motion and therefore can be used to label the trajectories.
For the example of two space-time dimensions, where there is only a constant electric field $F_{01}=E$,
solutions take the form 
\beqa x^0 &=& \frac1{E}\;(-P_1 \pm m \sinh \gamma(\tau) \;)\cr & &\cr 
x^1 &=& \frac1{E}\;(P_0 \pm m \cosh \gamma(\tau) \;)\;,\eeqa
where $\gamma(\tau)$ is arbitrary.

  The
reparametrization symmetry again leads to the mass shell constraint
(\ref{mschl}), only the momenta $p_\mu$ appearing there are not the canonical momenta.  Instead the
Poisson brackets (\ref{cpb}) are replaced by
\be 
\{x^\mu,p_\nu\} =\delta^\mu_\nu \qquad \{x^\mu,x^\nu\} =0 \qquad
\{p_\mu,p_\nu\} =-F_{\mu\nu}\label{ncpb}\; \ee
 $p_\mu$ do not have zero Poisson bracket with the  constraint
(\ref{mschl}), and thus are  not gauge invariant.   Nor
are they the conserved momenta $P_\mu$, which are related to $p_\mu$ by  
(\ref{capP}).  Since $\{P_\mu,p_\nu\}=0\;,$
it follows that  the conserved momenta are gauge invariant observables.
On the other hand, canonical momenta $\pi_\mu$ are constructed as follows
\be \pi_\mu = p_\mu +\frac12  F_{\mu\nu} x^\nu\;, \ee
and together with $j_{\mu\nu} = x_\mu \pi_\nu -x_\nu \pi_\mu$ generate
the Poincar\'e group.  However for nonvanishing fields the generators  are not gauge invariant observables.

In two space-time dimensions, a central extension $\widetilde
{ISO(1,1)}$  of the Poincar\'e
algebra  can be constructed.\cite{Cangemi:1992bj}  Moreover, its
generators are gauge invariant.  The translation generators are
$P_\mu$, and they have a central extension: \be \{P_\mu, P_\nu\} =E \epsilon_{\mu\nu}\;\label{cepwp}\ee
A gauge invariant  boost generator is 
\be K= \frac {E}2\; x^2
-\epsilon_{\mu\nu} x^\mu P^\nu \;,\ee and it leads to the usual
transformation properties for $P_\mu$ and $x^\mu$: 
  \beqa \{P_\mu, K\} &=& \epsilon_{\mu\nu} P^\nu \label{pbopwk}\\
\{x^\mu, K\} &=& \epsilon^{\mu\nu} x_\nu  \;,\qquad \label{pbbwx}\eeqa
From (\ref{pbopwk})  and (\ref{pbbwx}) it follows that $\{p_\mu, K\} =
\epsilon_{\mu\nu} p^\nu$, and hence that $K$ is gauge invariant.   $\widetilde
{ISO(1,1)}$ has the
 Casimir  \be
{\cal C}=  P^2 - 2 { E}\;K = (P_\mu -{  E}\;\epsilon_{\mu\nu}
x^\nu)^2 \;,
\ee
which from the mass shell constraint (\ref{mschl}) equals $ - m^2$.
We can therefore more generally add to the boost generator a term proportional to the Casimir,  preserving the Poisson brackets (\ref{cepwp}) and (\ref{pbopwk}):
\be  K \rightarrow K^{(\alpha)}=K + \frac\alpha{2E}\;{\cal C} = \frac
 \alpha {2E}\; P^2+ (\alpha - 1) \;K \;,\label{1p1bst}\ee
obtaining a  one parameter family of  $\widetilde
{ISO(1,1)}$ algebras.  Their generators are gauge invariant, and are  distinguished by
the Casimir, which has the value  \be
{\cal C}^{(\alpha)}=  P^2 - 2 { E}\;K^{(\alpha)} \approx (\alpha - 1) \;m^2\;,
\ee after using the mass shell constraint (\ref{mschl}).
However only for $\alpha = 0$, does $K^{(\alpha)}$
induce the  standard Lorentz boost  on space-time coordinates $x^\mu$
following from (\ref{pbbwx}).\footnote{  For the special case
  $\alpha= 1$,  the boost has the simple form $K^{(1)}=  \frac
 1 {2E}\; P^2$ and we can
define a new pair of gauge invariant space-time  coordinates $X^\mu$ which are just
the dual of $P_\mu$,
$$  X^\mu = \frac1E \; \epsilon^{\mu\nu} P_\nu\; $$  From
 $$ \{X^\mu, P_\nu\} =\delta^\mu_\nu \;,\qquad
\{X^\mu, K\} = \epsilon^{\mu\nu} X_\nu \label{pbxpj}$$ they undergo the usual two-dimensional Poincar\'e transformations.
Like $P_\mu$, they have nonvanishing Poisson brackets among themselves,
$ \{X^\mu, X^\nu\} = - E^{-1}\epsilon^{\mu\nu}$, and since they are reparametrization invariant merely serve to label the
orbits.}

Next  consider the gauge fixing.  We are again interested in a
nonstandard gauge
condition leading to the 
Dirac brackets (\ref{stnc}) and (\ref{dbxixj}), and so implying nontrivial commutation relations for the
space-time coordinates upon quantization.  This is accomplished for 
\be \phi_2 = x^0 + \theta^{0i} P_i - \tau\;\approx 0\;,\label{nsge} \ee 
$\theta^{0i}$ again being constants and $P_i$ being the gauge
invariant momenta.  It reduces to the previous gauge
condition (\ref{gfxng}) for
vanishing fields.  The Poisson bracket between
constraints $\phi_1$ and $\phi_2$ is again given by
(\ref{pboc}).  So we  recover
the previous 
Dirac brackets (\ref{stnc}) and (\ref{dbxixj}) between space-time
coordinates $x^\mu$, and commutation relations (\ref{stcrs}) upon quantization.  The remaining   Dirac brackets contain the interaction
with the constant field tensor.  The remaining   nonvanishing Dirac brackets are 
\beqa
 \{x^i,p_0\}_{DB}& =& N^i_{\;j} \frac {p_j}{p_0} \\
\{x^i,p_j\}_{DB}& =& N^i_{\;j} - \theta^{0i} F_{jk} \frac {p_k}{p_0}\\
\{p_0,p_i\}_{DB}& =&  F_{ij} \frac {p_j}{p_0}\label{thtynn}\\
\{p_i,p_j\}_{DB}& =&  - F_{ij}\;,\label{frty} \eeqa  
where $ N^i_{\;j}=\delta_{ij} -\theta^{0i}F_{0j}$.   It then follows
that \be\{P_0,P_i\}_{DB} =  F_{0i}\label{thnntn}\ee 

For the dynamics we again  write the Hamilton equations using
 Dirac brackets as in (\ref{heorps}).  Now we get that all the
 space-time coordinates  have an explicit $\tau$ dependence.   A convenient choice
for the
Hamiltonian is $P^0$, since it is the conserved energy.
 So setting $\phi_1$ strongly equal to zero, 
\be H=\sqrt{p_ip_i + m^2} - F_{0i} x^i\label{hamiltE}\ee  Since all
 $P_\mu$ should be  constants of the motion, from (\ref{thnntn}) we
 need that \be\frac{\partial
  P_0}{\partial\tau}= 0 \qquad
\frac{\partial
  P_i}{\partial\tau}=-F_{0i} \label{pdfps}\ee  Additional requirements on partial
derivatives come from demanding that  $\dot  \phi_a=\frac{\partial
  \phi_a}{\partial\tau}=0$, $a=1,2$, the dot again denoting a
total $\tau$ derivative.   They lead to 
\beqa p^\mu \frac{\partial
  p_\mu}{\partial\tau}&=& 0 \\  \frac{\partial
  x^0}{\partial\tau}&=& 1+F_{0k}\theta^{0k}\label{pdfxz} \eeqa 
A  solution consistent with (\ref{pdfps}-\ref{pdfxz}) is 
\beqa   \frac{\partial
  x^i}{\partial\tau}&=&-F_{0k}\theta^{0k}\frac {p_i}{p_0} \cr & &\cr
 \frac{\partial
  p_\mu}{\partial\tau}&=& F_{0k}\theta^{0k}F_{\mu\nu}\frac {p^\nu}{p_0}\;,
\eeqa and 
so all the phase space variables  $x^\mu$ and $p_\mu$  have
explicit $\tau$ dependence when the scalar product of $\theta^{0i}$
with  the electric
field $F_{0i}$ is not zero.
The resulting   Hamilton  equations of
motion are 
\beqa  \dot x^\mu & =& \{x^\mu, H\}_{DB}+\frac{\partial
  x^\mu}{\partial\tau}\; =\;-\frac {p^\mu}{p_0} \cr & &\cr
 \dot p_\mu & =& \{p_\mu, H\}_{DB}+\frac{\partial
  p_\mu}{\partial\tau}\; =\;F_{\mu\nu}\frac {p^\nu}{p_0}\;,
\label{eomE1} \eeqa  which agrees with (\ref{cpivemf}).  As in the
 free case,  $ \dot x^0 =1$, and  the commutative and noncommutative clock, as measured by
 $\tau$ and $x^0$, respectively, run at the same rate.

 Assuming $[ N^i_{\;j}]$ to be a nonsingular matrix
  ($F_{0k}\theta^{0k}\ne 1$), the noncommutativity of the space-time coordinates following from
  (\ref{stnc}) and (\ref{dbxixj}) can again be removed by  a trivial
  coordinate transformation.   It now takes the form
\beqa  x^i&\rightarrow & q^i = N^{-1\;}{}^i_{\;j}\; [x^j + \theta^{0j}
  p_0]\cr
& &\cr
 x^0&\rightarrow & q^0 = x^0 + \theta^{0i} P_i =\tau\label{trsfmxtqf} \eeqa 
 The coordinates  $q^\mu$ once again satisfy the standard gauge, and
  have its  associated Dirac brackets  \beqa 
  \{q^i,p_0\}_{DB}& =& \frac {p_i}{p_0}\label{sboqipz}\\
\{q^i,p_j\}_{DB}& =& \delta_{ij}\;,\label{sboqipj} \eeqa along with
  $\{q^\mu,q^\nu\}_{DB}= \{q^0,p_\nu\}_{DB}=0$, (\ref{thtynn}) and
  (\ref{frty}). 
  Conversely, we can start with the standard  gauge, and 
obtain  the  gauge (\ref{nsge}) by applying the inverse of transformation (\ref{trsfmxtqf}),
\beqa x^0 &=& \frac{ q^0 -\theta^{0i}(p_i + F_{ij} q^j)}{1-\theta^{0k}
  F_{0k}}\cr &&\cr
 x^i &=& N^i_{\;j} q^j - \theta^{0i} p_0\; \eeqa
   So once again both gauges give rise to the same
  Poisson structure and resulting quantum commutation relations.
Concerning the dynamics, the natural Hamiltonian in the standard gauge would be
  (\ref{hamiltE}) with noncommuting coordinates  $x^i$ replaced by
  commuting ones $q^i$: \be  H_{0}=\sqrt{p_ip_i + m^2} - F_{0i} q^i
  \ee  It now represents the conserved energy, and
yields the same equations of motion  as (\ref{eomE1}).  (Now $q^i$ and
  $p_\mu$ have no explicit $\tau$ dependence.)

\section{Other Interactions}
\setcounter{equation}{0}

For arbitrary interactions 
 there is no longer,  in general, a
conserved momenta.  The latter was used previously in writing
the  gauge condition (\ref{nsge}), and it led to the simple
commutation relations (\ref{stcrs}) between the space and time
coordinates.  It also implied that  the commutative and noncommutative clock, as measured by
 $\tau$ and $ x^0$, respectively, run at the same rate,
 i.e. $\frac{d{\bf x}^0}{d\tau}=1$.  For more general systems,
these results get altered.  Moreover, one can even have  $\frac{d{\bf
    x}^0}{d\tau}<0$ implying time reversal in transforming from time
$\tau$ to time ${\bf x}^0$.

\subsection{Coupling to an Arbitrary Electromagnetic Field}

 The first example is the case of a relativistic particle coupled to an arbitrary
electromagnetic field.   As before the action is reparametrization
invariant.  Here we replace (\ref{cfcplng}) by  \be
S_F =-\int d\tau \;A_\mu(x)\; \dot x^\mu \;\;,
\ee with the resulting equations of motion (\ref{cpivemf}), where $
F_{\mu\nu}=\partial_\mu A_\nu- \partial_\nu A_\mu$ is not in general constant.
  The mass shell constraint
(\ref{mschl}) and  
Poisson brackets (\ref{ncpb}) once again follow.  If for the gauge
constraint one takes (\ref{gfxng}), then (\ref{pboc}) gets replaced by
\be \{\phi_1,\phi_2\} = 2( p_0 +\theta^{0i} F_{i\mu} p^\mu)\;, \ee
leading to a rather complicated Dirac bracket
between the space and time coordinates
\be  \{x^0,x^i\}_{DB} = \frac{\theta^{0i}}{1+\theta^{0i} F_{i\mu}
 \frac{ p^\mu}{p_0}}\;,
\ee as opposed to the result obtained previously (\ref{stnc}). Moreover, demanding
 that $\dot\phi_2=0$ now gives the complicated result
\be \dot x^0 =\frac {1+ \theta^{0i}F_{ij}\dot
  x^j}{1+\theta^{0k}F_{0k}}\;,\ee
as opposed to $\dot x_0=1$.

An alternative approach is to start with the standard gauge
$\phi_2= q^0-\tau\approx 0$ (here we denote the space-time coordinates by $q^\mu$), and simply define a noncommutative time, using
for example (\ref{bfxz}).  The nonvanishing  Dirac brackets in the
standard gauge are again given by  (\ref{thtynn}),
  (\ref{frty}), (\ref{sboqipz}) and (\ref{sboqipj}).   The dynamics in
  the standard gauge is recovered for the Hamiltonian
\be  H_{0}=\sqrt{p_ip_i + m^2} +A_0(q) \;,  \ee
along with \be
\frac {\partial p_i}{\partial\tau} = \partial_0A_i\qquad  \quad
\frac {\partial p_0}{\partial\tau} = \partial_0A_i\frac{p^i}{p_0}\qquad\quad  
\frac {\partial q^0}{\partial\tau} = 1\;,\ee which is  consistent with
the conditions $\dot \phi_a = \frac {\phi_a}{\partial\tau}=0,\;a=1,2$.
Now define $x^0=q^0-\theta^{0i}p_i$ to obtain the familiar Dirac brackets
\be \{x^0,q^i\}_{DB} = \theta^{0i}\ee  A feature shared with the
previous approach is that $\dot x^0 \ne 1$.  Now
\be \dot x^0 = 1 - \theta^{0i}F_{0i} +\theta^{0i}F_{ij} \dot q^j \ee
Since this approach differs from the previous one only by a gauge
choice, the dynamics in the two cases must be identical.  The
difference between the two approaches is in how the time variable
$x^0$ is defined.  For both definitions $\dot x^0\ne 1$, and even
allows for the possibility of time reversal in going from time as
measured by $\tau$ to time as measured by $x^0$.

\subsection{Conservative System}

In all the previous examples, a
 noncommutative time resulted either  from a gauge choice or by a
 redefinition  of coordinates.  In sections two and three these approaches
 were equivalent, while in the above example one ends up with
 different definitions of the noncommutative time $x^0$.  In systems with no time reparametrization symmetry,
 one can adapt the second approach.  So once again by defining
 (\ref{bfxz}) and assuming the commutation relations (\ref{ccrqp}),
 the result (\ref{stcrs})
  follows.   Applying this to a  nonrelativistic
 conservative system described by Hamiltonian \be H_{0}= \frac{p_i^2}{2m}
 + V(q^i)\;,\label{hsgv}\ee one gets
\be \dot x^0=1+\theta^{0i}\frac{\partial V}{\partial
 q^i}\;, \label{dtdtsg}\ee where $H_{0}$ generates evolution in
 $\tau$.   If there are trajectories
  for which  $1+\theta^{0i}\frac{\partial V}{\partial
 q^i}<0$ ,  we then get a time reversal upon applying (\ref{bfxz}).

In the above we looked at  replacing the commuting time with its noncommuting
  counterpart,  using
 (\ref{bfxz}).  One can instead  make the analogous replacement of the spatial
  coordinate.     For the free particle this corresponded to the inverse of
 (\ref{mpxiqi}), or
\be  q^i\rightarrow  x^i = q^i - \theta^{0i} H\label{rlcqx} \ee  One can try
  repeating this for an interacting system, with  $H$ representing the
  resulting Hamiltonian for the system generating evolution in some
  new time variable, which we denote by  $\tau'$.  The generalizations  of (\ref{dbxixj}) and (\ref{ten})
 are then \beqa
 \{x^i,x^j\}& =&\theta^{0i} \frac{d x^j}{d\tau'} -\theta^{0j}  \frac{d
 x^i}{d\tau'}\cr & &\cr
\{x^i,p_j\}& =& \delta_{ij} +\theta^{0i} \frac{d p_j}{d\tau'}\;, \label{fftytw}\eeqa
So   starting from the  nonrelativistic
 conservative Hamiltonian (\ref{hsgv}), we would get \be H=
  \frac{p_i^2}{2m} + V(x^i)\;,  \label{fftyfr}\ee upon making the
  replacement  (\ref{rlcqx}).
The  Hamilton equations of motion resulting from (\ref{fftytw}) and (\ref{fftyfr}) can be written
\beqa\biggl(1+\theta^{0j}\frac{\partial V}{\partial x^j}\biggr)
 \;\frac{ d  x^i}{d\tau'}  &=& \frac {p_i} m \cr & &\cr
 \biggl(1+\theta^{0j}\frac{\partial V}{\partial x^j}\biggr) \;\frac {
 d p_i}{d\tau'}&= &- \frac{\partial V}{\partial x^i} \eeqa  
Provided $1+\theta^{0j}\frac{\partial V}{\partial x^j}>0$, the  associated classical trajectories are identical to those
  generated from the standard Hamiltonian (\ref{hsgv})
  after again  performing a reparametrization
\be \frac{d\tau'}{d\tau}=1+\theta^{0i}\frac{\partial V}{\partial
 x^i} \ee  We thus arrive at the same Jacobian factor as
  in (\ref{dtdtsg}).  Unlike in the previous paragraph, here both
  `times' are associated with c-numbers.   As before, 
 if there are trajectories
  for which  $1+\theta^{0i}\frac{\partial V}{\partial
 x^i}<0$ ,  we then get a time reversal upon going from $\tau$ to
  $\tau'$. 

\bigskip
\bigskip

{\parindent 0cm{\bf Acknowledgement}}                                         
 
We are very grateful to A.P. Balachandran, G. Bimonte and R. Sorkin for
discussions.  This work was supported in part by the joint NSF-CONACyT grant
E120.0462/2000 and  DOE grant   DE-FG02-85ER40231.

\end{document}